\documentclass[twocolumn,fullpaper]{jpsj3}

\usepackage{graphicx}
\usepackage{dcolumn}
\usepackage{bm}
\usepackage{color}
\usepackage{ulem}
\usepackage{booktabs}
\usepackage[switch]{lineno} 

\begin{document}

\title{Intrinsic Magnetic Excitations and Heavy-Fermion Formation in the Frustrated Mn Pyrochlore System YMn$_{2+\delta}$Zn$_{20-x}X_x$ ($X$ = In and Al) Revealed by Nuclear Magnetic Resonance and Nuclear Quadrupole Resonance Measurements}

\author{Shunsaku~Kitagawa$^{1,}$\thanks{E-mail address: kitagawa.shunsaku.8u@kyoto-u.ac.jp}, 
Kenji~Ishida$^{1}$,
Yoshihiko~Okamoto$^{2,}$, and
Zenji~Hiroi$^{2}$
}

\inst{$^1$Department of Physics, Graduate School of Science, Kyoto University, Kyoto 606-8502, Japan \\
$^2$Institute for Solid State Physics, University of Tokyo, Kashiwa 5-1-5, Chiba 277-8581, Japan 
}

\date{\today}

\abst{
We performed nuclear magnetic resonance (NMR) and nuclear quadrupole resonance (NQR) measurements to investigate the microscopic electronic states of the $d$-electron heavy-fermion candidates $\mathrm{YMn_{2+\delta}Zn_{20-x}In_x}$ and $\mathrm{YMn_{2+\delta}Zn_{20-x}Al_x}$. 
In these compounds, magnetic fluctuations of the Mn pyrochlore lattice are expected to play an important role in heavy-fermion formation; however, excess Mn atoms complicate the interpretation of the physical properties.
Our spectral analysis reveals that In substitution exhibits much higher site selectivity and introduces significantly less disorder in local structure than Al substitution.
The temperature dependence of the nuclear spin-lattice relaxation rate divided by temperature $1/T_1T$ measured by $^{55}$Mn-NQR shows a clear enhancement at low temperatures, indicating the development of low-energy excitations associated with heavy-fermion formation.
However, its absolute magnitude is approximately 20 times smaller than that in the related compound YMn$_2$, which hosts stronger antiferromagnetic correlations, indicating that the magnetic interactions are substantially weakened by the enlarged Mn-Mn distance.
These results demonstrate that the heavy-fermion state in this system arises from the Mn pyrochlore network and is more closely associated with frustration-induced magnetic excitations with low energy than with conventional antiferromagnetic quantum-critical fluctuations.
}

\maketitle

\section{Introduction}
\begin{figure}[!tb]
\centering
\includegraphics[width=8.3cm,clip]{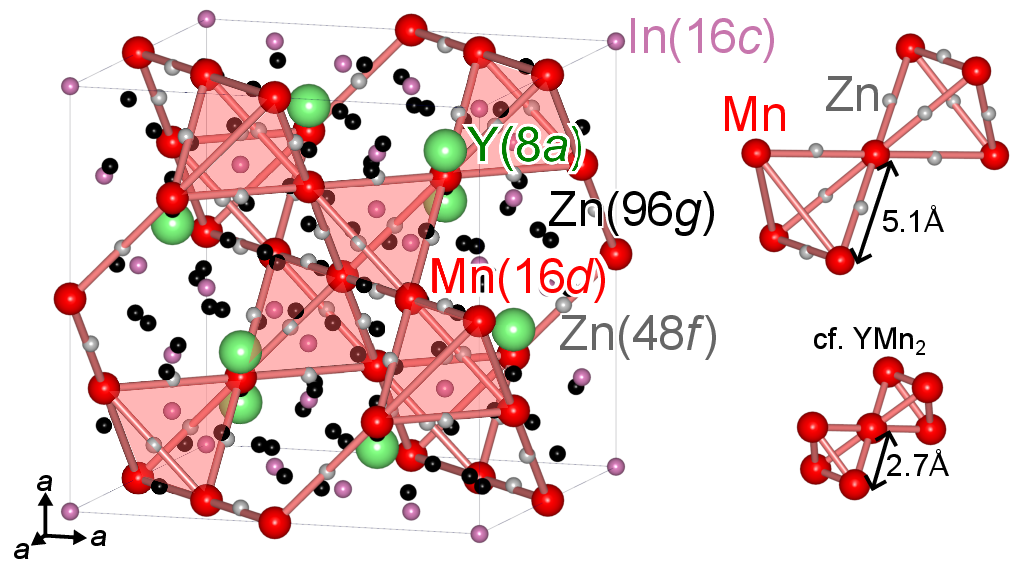}
\caption{
(Color online) Crystal structure of  In-substituted $\mathrm{YMn_2Zn_{20}}$ drawn by VESTA~\cite{K.Momma_JAC_2011}.
A box indicates the unit cell.
}
\label{Fig.1}
\end{figure}

In recent years, the $\mathrm{AB_2C_{20}}$ ($\mathrm{A}$ = rare earth, $\mathrm{B}$ = transition metal, $\mathrm{C}$ = $\mathrm{Zn, Al}$) family of cage compounds crystallizing in the cubic $\mathrm{CeCr_2Al_{20}}$-type structure (space group $Fd\bar{3}m$, $O_{h}^{7}$, and No. 227) has attracted attention as a novel platform for studying strongly correlated electron systems\cite{T.Nasch_ZNaturforsch_1997,T.Onimaru_JPSJ_2016}.
In this crystal structure, as shown in Fig.~\ref{Fig.1}, the $\mathrm{A}$ atoms are encapsulated within giant polyhedral cages formed by 16 $\mathrm{C}$ atoms, creating a diamond lattice.
Meanwhile, the transition metal $\mathrm{B}$ atoms form a network of corner-sharing tetrahedra, namely, a pyrochlore lattice, which is the most prominent feature of this system.

This structure induces a variety of physical phenomena.
For instance, the low-energy anharmonic oscillation, or rattling, of the $\mathrm{A}$ atoms inside the cage causes a dramatic decrease in thermal conductivity due to phonon scattering, enhancing the material's potential for thermoelectric applications\cite{Keppens_Nature_1998,Sales_Science_1996,K.Wakiya_PRB_2016,K.Wei_SciAdv_2019}.
Furthermore, regarding the electronic states, $\mathrm{YbCo_2Zn_{20}}$ exhibits an extreme heavy fermion state with an electronic specific heat coefficient reaching $\gamma \approx 8$ $\mathrm{J}$ $\mathrm{K^{-2}}$ $\mathrm{mol^{-1}}$\cite{Torikachvili_PNAS_2007}.
Numerous unique physical properties originating from the $c$-$f$ hybridization have also been reported\cite{A.Sakai_JPSJ_2011,E.D.Bauer_PRB_2008}.

Typically, heavy-fermion systems are formed in rare-earth and actinide compounds possessing 4$f$ or 5$f$ electrons via the Kondo effect\cite{G.R.Stewart_RMP_1984,P.A.Lee_comments_1986,S.Kitagawa_JPSJ_2022,S.Kitagawa_JPSJ_2025}.
However, in $d$-electron systems, when geometrical frustration—such as that found in a pyrochlore lattice—suppresses antiferromagnetic ordering, the residual strong spin fluctuations can increase the effective mass of electrons, leading to the realization of a heavy-fermion state\cite{R.Moessner_PRB_1998,S.Tanaka_JPSJ_2009,C.Kacroix_JPSJ_2010}.
Representative examples include the spinel oxide $\mathrm{LiV_2O_4}$\cite{S.Kondo_PRL_1997,C.Urano_PRL_2000} and the Laves phase compound $\mathrm{(Y,Sc)Mn_2}$\cite{M.Shiga_PhysicaBC_1988,R.Ballou_PRL_1996}.
$\mathrm{YMn_2Zn_{20}}$ was discovered as a candidate for a $d$-electron heavy fermion material\cite{Y.Okamoto_JPSJ_2010}.
Since $\mathrm{Y}$ is nonmagnetic, the magnetism of this system is exclusively governed by the $\mathrm{Mn}$ 3$d$ electrons.
Although synthesizing pure $\mathrm{YMn_2Zn_{20}}$ is difficult, partially substituting the $\mathrm{Zn}$ sites with $\mathrm{In}$ or $\mathrm{Al}$ stabilizes the structure, enabling single-crystal growth\cite{Y.Okamoto_JPSJ_2010,Y.Okamoto_JSSC_2012}.
Particularly in samples with low $\mathrm{In}$-substitution concentrations, a specific heat coefficient of $\gamma \approx 280$ $\mathrm{mJ}$ $\mathrm{K^{-2}}$ $\mathrm{mol^{-1}}$ has been observed\cite{Y.Okamoto_JPSJ_2010}, which is double that of Y$_{0.97}$Sc$_{0.03}$Mn$_2$($\approx 140$ $\mathrm{mJ}$ $\mathrm{K^{-2}}$ $\mathrm{mol^{-1}}$)\cite{H.Imai_JPSJ_1995}.
The comparison between the two systems provides an opportunity to discuss the mechanism of the heavy-fermion state.

\begin{table*}[h]
\centering
\caption{Crystallographic parameters for In-substituted $\mathrm{YMn_{2.11}Zn_{17.53}In_{2.36}}$ determined by single-crystal XRD measurements\cite{Y.Okamoto_JSSC_2012}.}
\begin{tabular}{llllll}
\hline
& Wyckoff & Site & $x$ & $y$ & $z$ \\
& position & symmetry & & & \\
\hline
Y   & $8a$  & \={4}3m & $1/8$        & $1/8$        & $1/8$        \\
Mn  & $16d$ & .\={3}m & $1/2$        & $1/2$        & $1/2$        \\
In  & $16c$ & .\={3}m & $0$          & $0$          & $0$          \\
Zn2 & $48f$ & 2.mm & $0.49092(2)$ & $1/8$        & $1/8$        \\
Zn1 & $96g$ & ..m & $0.05703(1)$ & $0.05703(1)$ & $0.32842(2)$ \\
\hline
\end{tabular}
\label{tab1}
\end{table*}

However, the fundamental understanding of the physics in this system is hindered by chemical compositional inhomogeneity.
Inductively-coupled plasma atomic-emission (ICP-AES) and single-crystal X-ray diffraction (XRD) measurements have revealed excess Mn, namely that Mn occupies not only the intrinsic 16$d$ sites but also Zn sites as the In content increases to stabilize the structure.
These excess $\mathrm{Mn}$ atoms possess a large magnetic moment close to a localized spin $S = 5/2$, causing spin-glass-like behavior at low temperatures\cite{Y.Okamoto_JPSJ_2010}.
Therefore, distinguishing whether the observed mass enhancement is due to itinerant $\mathrm{Mn}$ electrons on the pure pyrochlore lattice or an impurity effect from excess $\mathrm{Mn}$ is a critical issue.

In this study, we aim to verify the site-selective structural stabilization and elucidate the intrinsic magnetic excitations of the Mn pyrochlore network in In- and Al-substituted YMn$_{2}$Zn$_{20}$ systems by using the nuclei of Mn and the substituted elements as microscopic nuclear magnetic resonance (NMR)/nuclear quadrupole resonance (NQR) probes.
Because NMR is highly sensitive to local electronic and structural environments\cite{S.Kitagawa_PRB_2018,S.Kitagawa_PRB_2024}, it enables us to clearly distinguish the intrinsic signals of itinerant Mn atoms on the 16$d$ pyrochlore lattice from those of the localized excess Mn impurities. 
Through this site-selective local probe, we reveal that the In-substituted system exhibits significantly higher crystalline order and site selectivity than the Al-substituted system.
Furthermore, by exclusively isolating the NMR/NQR responses from the intrinsic pyrochlore network, we successfully observed a clear enhancement of $1/T_1T$ at low temperatures.
This provides microscopic evidence for enhanced low-energy excitations intrinsic to the Mn pyrochlore lattice, which is considered to origin of the heavy-fermion formation in this geometrically frustrated system.

\section{Experimental}
Powdered polycrystalline samples, prepared by crushing crystals grown via a melt-growth method, were used in this study.
For comparative analysis, we selected two samples: $\mathrm{In}$-substituted system $\mathrm{YMn_{2.11}Zn_{17.53}In_{2.36}}$, and $\mathrm{Al}$-substituted system $\mathrm{YMn_{2.06}Zn_{12.23}Al_{7.71}}$.
The chemical compositions of the In- and Al-substituted samples were determined by ICP-AES spectroscopy and single-crystal XRD measurements.
Crystallographic parameters for the In-substituted system determined by single-crystal XRD measurements\cite{Y.Okamoto_JSSC_2012} are listed in Table I.
$^{55}\mathrm{Mn}$ (nuclear spin $I = 5/2$, gyromagnetic ratio $^{55}\gamma/2\pi = 10.554$~MHz/T, nuclear electric quadrupole moment $^{55}Q = 0.330 \times 10^{-28}$ m$^2$), $^{115}\mathrm{In}$ ($I = 9/2$, $^{115}\gamma/2\pi = 9.3296$~MHz/T, $^{115}Q = 0.770 \times 10^{-28}$ m$^2$), and $^{27}\mathrm{Al}$ ($I = 5/2$, $^{27}\gamma/2\pi = 11.094$~MHz/T, $^{27}Q = 0.1466 \times 10^{-28}$ m$^2$)-NMR/NQR measurements were performed with a conventional spin-echo technique~\cite{R.K.Harris_2001,N.J.Stone_Q_2016}.
The NMR(NQR) spectra as a function of the magnetic field (frequency) were obtained using the Fourier transform of a spin–echo signal observed after a radio-frequency pulse sequence at a fixed frequency (zero field).
A nuclear spin-lattice relaxation rate $1/T_1$ was evaluated by fitting the relaxation curve of the nuclear magnetization after the saturation $M(t)$ to a theoretical function for the nuclear spin $I = 5/2$ and that for $I = 9/2$ as follows,
\begin{align}
M(t) = M(\infty)\left\{ 1 - \left[\frac{3}{28}\exp\left( -\frac{3t}{T_1}\right)+\frac{25}{28}\exp\left( -\frac{10t}{T_1}\right)  \right] \right\}& \notag\\
\text{for $^{55}$Mn-NQR $\left(\pm\frac{1}{2} \leftrightarrow \pm\frac{3}{2}\right)$}& \notag,\\
M(t) = M(\infty)\left\{ 1 - \left[\frac{3}{7}\exp\left( -\frac{3t}{T_1}\right)+\frac{4}{7}\exp\left( -\frac{10t}{T_1}\right)  \right] \right\}& \notag\\
\text{for $^{55}$Mn-NQR $\left(\pm\frac{3}{2} \leftrightarrow \pm\frac{5}{2}\right)$}& \notag,\\
M(t) = M(\infty)\bigg\{ 1 - \bigg[\frac{1}{33}\exp\bigg( -\frac{3t}{T_1}\bigg)+\frac{20}{143}\exp\bigg( -\frac{10t}{T_1}\bigg)&\notag\\
+ \frac{4}{165}\exp\bigg( -\frac{21t}{T_1}\bigg)+\frac{576}{715}\exp\bigg( -\frac{36t}{T_1}\bigg) \bigg] \bigg\}& \notag\\
\text{for $^{115}$In-NQR $\left(\pm\frac{3}{2} \leftrightarrow \pm\frac{5}{2}\right)$}& \notag.
\end{align}
We measured $1/T_1$ for the $^{55}$Mn-NQR by using the $\pm 3/2 \leftrightarrow \pm 5/2$ transition above 1.5 K, while the $\pm 1/2 \leftrightarrow \pm 3/2$ transition was used below 1.5 K.
Temperature control down to 1.5~K and 0.06~K was achieved using a $^4\mathrm{He}$ cryostat and a $^3$He–$^4$He dilution refrigerator, respectively.

\begin{figure*}[!tb]
\centering
\includegraphics[width=16cm,clip]{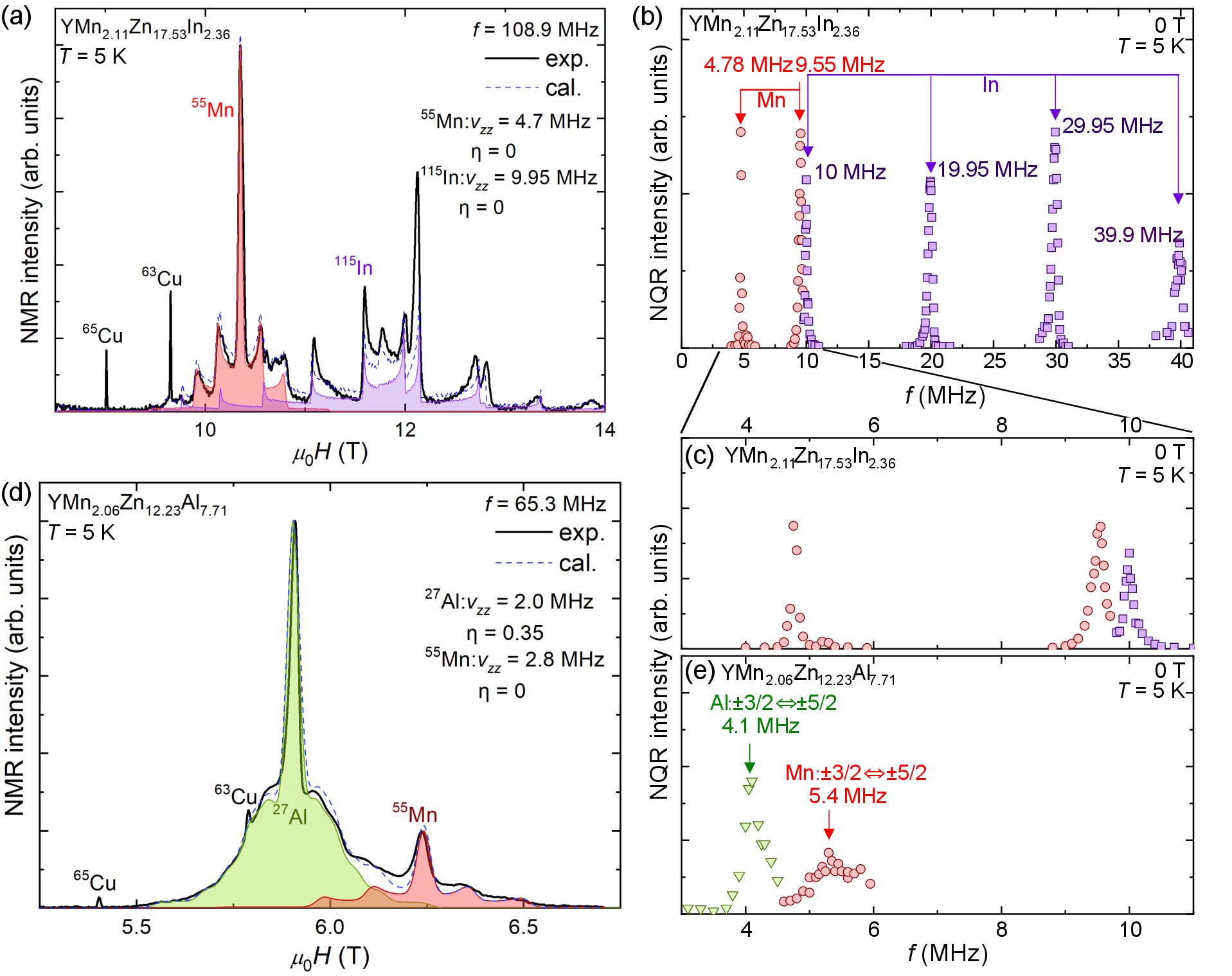}
\caption{
(Color online) The results of NMR and NQR spectra measurements. 
Field-swept NMR spectrum with 108.9 MHz (a) and frequency-swept NQR spectrum (b) of $\mathrm{YMn_{2.11}Zn_{17.53}In_{2.36}}$ measured at 5 K.
(c) Expanded view of the NQR spectrum in the low-frequency region.
Field-swept NMR spectrum with 65.3 MHz (d) and frequency-swept NQR spectrum (e) of $\mathrm{YMn_{2.06}Zn_{12.23}Al_{7.71}}$ measured at 5 K.
$^{63/65}$Cu-NMR signal comes from NMR coil.
Numerical simulation of randomly oriented component is also plotted.
The NQR parameters of the simulation are shown in the figure.
}
\label{Fig.2}
\end{figure*}

\section{Results and Discussion}
\subsection{Local structure of the YMn$_2$Zn$_{20}$-based system}
To examine the effect of substitution on the local structure of YMn$_2$Zn$_{20}$-based system, we performed NMR/NQR measurements.
Figure~\ref{Fig.2} (a) shows the NMR spectrum of $\mathrm{YMn_{2.11}Zn_{17.53}In_{2.36}}$ at 5~K.
The NMR spectrum consists of several well-defined peaks, indicating a homogeneous electric and magnetic environment.

When $I \ge 1$, a nucleus has an electric quadrupole moment $eQ$ as well as a magnetic dipole moment, and thus, the degeneracy of nuclear-energy levels is lifted even at zero magnetic field due to the interaction between $eQ$ and the electric field gradient (EFG). 
In addition, Kramers degeneracy is lifted by applying a magnetic field.
The total effective Hamiltonian can be written as,
\begin{align}
\mathcal{H} &= -\frac{\gamma}{2\pi} h (1 + K)\bm{I} \cdot \bm{H}  \notag \\
            &+ \frac{h \nu_{zz}}{6}\left\{(3I_z^2-I^2)+\frac{1}{2}\eta(I_+^2+I_-^2)\right\},
\label{eq.H}
\end{align}
where $\gamma$ is nuclear gyromagnetic ratio, $h$ is Planck's constant, $K$ is the Knight shift, $\bm{H}$ is an external field, $\nu_{zz}$ is the quadrupole frequency along the principal axis of the EFG, and is defined as $\nu_{zz} \equiv 3e^2qQ/2I(2I-1)$ with $eq = V_{zz}$, and $\eta$ is an asymmetry parameter of the EFG expressed as $(V_{xx}-V_{yy})/ V_{zz}$ with $V_{\alpha \alpha}$, which is the second derivative of the electric potential $V$ and is the EFG along the $\alpha$ direction ($\alpha = x,y,z$).
To obtain a resonance field, we diagonalize Eq.\eqref{eq.H}.
As the NMR resonance condition depends on the angle between the principal axis of the EFG and magnetic field direction, the sum of the spectrum for all angles is observed in the case of powder samples.

As shown in Fig.~\ref{Fig.2} (a), the experimental NMR spectrum is well reproduced by $\nu_{zz} = 4.7$ MHz and $\eta = 0$ for $^{55}$Mn nucleus and $\nu_{zz} = 9.95$ MHz and $\eta = 0$ for $^{115}$In nucleus.
The small differences are due to partial orientation toward the [1,1,1] direction, as indicated by the dotted line in Fig.~\ref{Fig.2} (a).
The obtained NQR parameters were confirmed by the NQR spectrum shown in Figs.~\ref{Fig.2} (b) and \ref{Fig.2} (c).
Reflecting the threefold rotational symmetry, the Mn site (16$d$) has an asymmetry parameter of  $\eta = 0$.
The relatively narrow spectral linewidth indicates that the Mn sites forming the pyrochlore lattice are maintained in a crystallographically homogeneous environment.
The In site (16$c$) also has $\eta = 0$. 
These results support the XRD results that In substitution occurs selectively at the Zn 16$c$ sites without compromising the uniformity of the Mn pyrochlore network\cite{Y.Okamoto_JSSC_2012}.

In contrast to the In-substituted sample, the NMR spectrum of the Al-substituted sample exhibited significant broadening, as shown in Fig.~\ref{Fig.2} (d).
The experimental spectrum can be reproduced by $\nu_{zz} = 2.8$ MHz and $\eta = 0$ for $^{55}$Mn nucleus and $\nu_{zz} = 2.0$ MHz and $\eta = 0.35$ for $^{27}$Al nucleus, together with a large distribution of $\nu_{zz}$.
The corresponding NQR spectrum is shown in Fig.~\ref{Fig.2} (e). 
Owing to the limitations of the experimental conditions, however, the $\pm 1/2 \leftrightarrow \pm 3/2$ transition could not be detected.
Broad $^{55}$Mn-NMR/NQR spectra indicate a large spatial distribution in the EFG ($\nu_{zz}$ and $\eta$) around the Mn nuclei.
This disorder is attributed to the random distribution of Al atoms over multiple Zn sites ($16c$, $48f$, and $96g$) due to the relatively large amount of Al substitution and the similarity in atomic radii between Al and Zn, as reported in the previous study\cite{Y.Okamoto_JSSC_2012}.
According to Ref.~22, Al atoms preferentially occupy the $48f$ site over the $16c$ site.
However, the Al substitution for Zn atoms occurs more randomly than the In substitution, resulting in significant structural disorder.

\begin{figure}[!tb]
\centering
\includegraphics[width=8.3cm,clip]{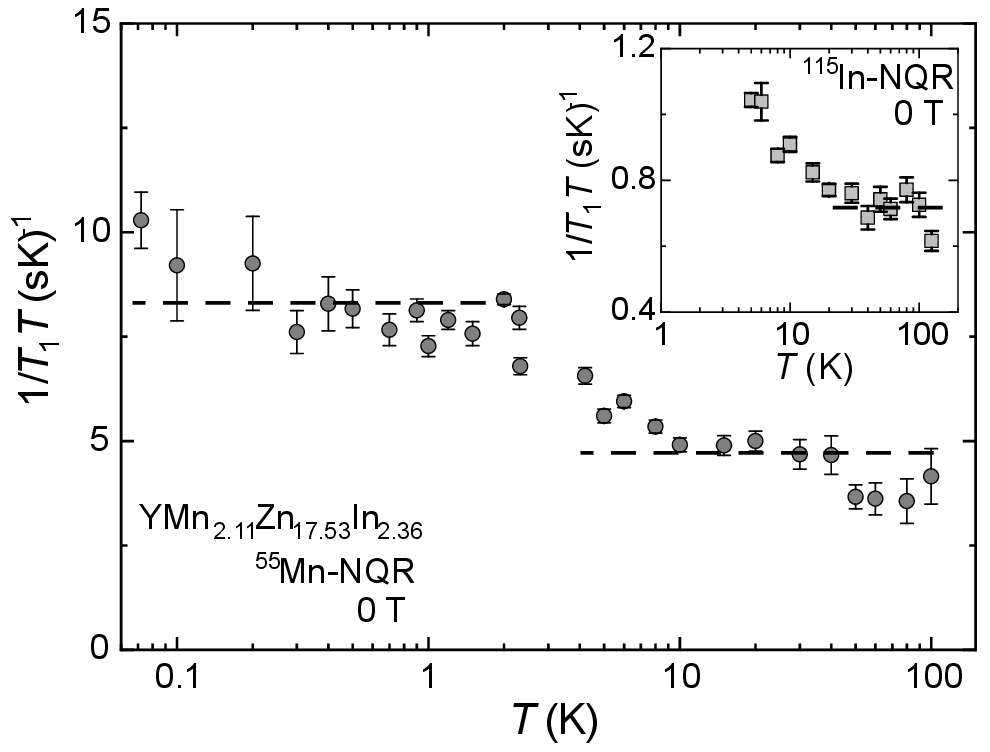}
\caption{
Temperature dependence of $1/T_1T$ of $^{55}$Mn-NQR in $\mathrm{YMn_{2.11}Zn_{17.53}In_{2.36}}$ measured at zero field.
The dotted lines are guides for eyes.
(Inset) Temperature dependence of $1/T_1T$ of $^{115}$In-NQR in $\mathrm{YMn_{2.11}Zn_{17.53}In_{2.36}}$ measured at zero field.
}
\label{Fig.3}
\end{figure}

\begin{figure}[!tb]
\centering
\includegraphics[width=8.3cm,clip]{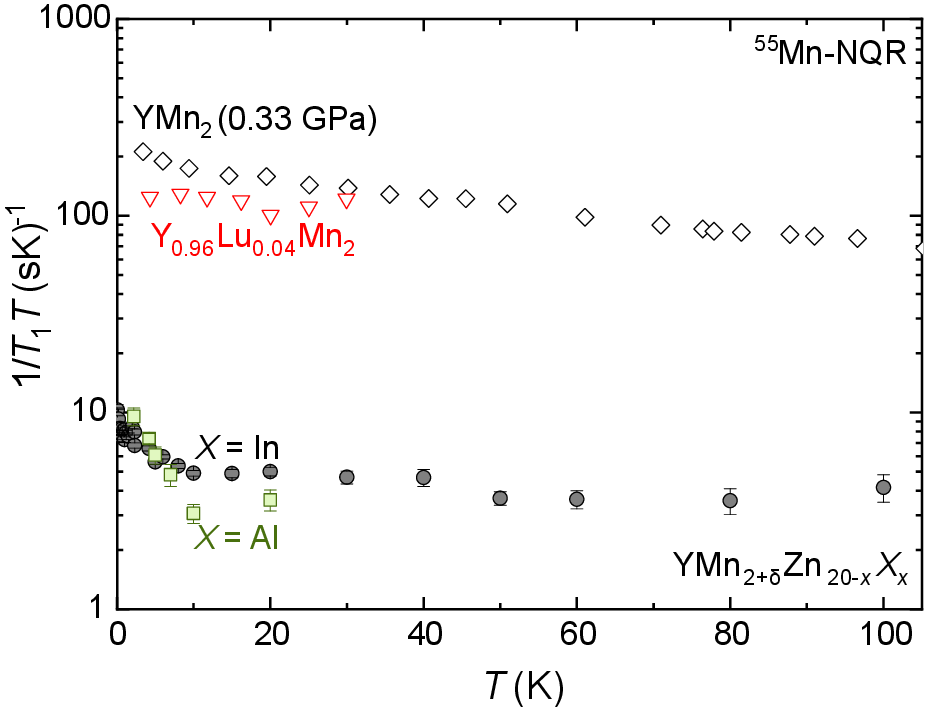}
\caption{
(Color online) The comparison of the temperature dependence of $1/T_1T$ of $^{55}$Mn-NQR in the YMn$_2$Zn$_{20}$-based and the YMn$_2$-based systems\cite{Z.Guo-zing_PRB_1999,M.Shiga_JPSJsupA_2000}.
}
\label{Fig.4}
\end{figure}

\subsection{Magnetic dynamics}
To investigate the low-energy magnetic dynamics, we measured the temperature dependence of $1/T_1T$ of $^{55}$Mn-NQR in $\mathrm{YMn_{2.11}Zn_{17.53}In_{2.36}}$, as shown in Fig.~\ref{Fig.3}.
In the high-temperature region, $1/T_1T$ remained approximately constant, indicating a normal metallic state (Korringa relation).
As the temperature decreased, a distinct increase in $1/T_1T$ was observed below 10~K, which is consistent with the enhancement of the electronic specific heat coefficient $\gamma$\cite{Y.Okamoto_JPSJ_2010}.
This provides microscopic evidence for enhanced low-energy excitations, which is consistent with heavy-fermion formation.

A similar enhancement in $1/T_1T$ was also observed at the $^{115}$In site (inset of Fig.~\ref{Fig.3}) as well as in the Al-substituted $\mathrm{YMn_{2.06}Zn_{12.23}Al_{7.71}}$ (Fig.~\ref{Fig.4}).
The difference in the absolute values is attributed to the difference in the hyperfine coupling constants $A_{\text{hf}}$.
The similar temperature dependence indicates that the enhancement is an intrinsic property of the Mn pyrochlore network rather than a site-specific anomaly.

It is worth noting that an anomaly around 10 K has also been reported in muon spin relaxation ($\mu$SR) measurements \cite{M.Miyazaki_JPCS_2014}. 
While the $\mu$SR study suggested quasi-one-dimensional spin dynamics below this temperature, such details are not directly resolved within the present NQR frequency window.
Instead, $1/T_1T$ tends to level off or shows weak increase at the lowest temperatures, which is consistent with the formation of a Fermi-liquid ground state.

Compared with the Laves-phase compound $\mathrm{YMn_2}$, the absolute value of $1/T_1T$ in the present system is approximately 20 times smaller as shown in Fig.~\ref{Fig.4}.
In general, such a difference originates from a reduction in either $A_{\text{hf}}$ or the amplitude of magnetic fluctuations.
Although a quantitative estimation of $A_{\text{hf}}$ via a Knight shift-$\chi$ plot is difficult due to the contribution of excess Mn impurities to the bulk susceptibility, the significantly larger $1/T_1T$ at the Mn site compared to the In site suggests that $A_{\text{hf}}$ for the $^{55}$Mn site would be  comparable to that of $\mathrm{YMn_2}$\cite{Z.Guo-zing_PRB_1999}.
Therefore, the smaller $1/T_1T$ directly reflects a substantial suppression of magnetic fluctuations in the present system.
This suppression can be qualitatively understood by the significant expansion of the Mn-Mn distance $r_\mathrm{Mn-Mn}$ from 2.7~\AA~to 5.1~\AA, as shown in Fig.~\ref{Fig.1}.
As the exchange interaction $J$ between Mn 3$d$ orbitals depends sensitively on the spatial overlap of wavefunctions, this nearly twofold increase in distance leads to a drastic reduction in $J$.
From the viewpoint of the dynamical susceptibility $\chi''(q, \omega)$, the weakened $J$ suppresses the strong antiferromagnetic correlations and induces the development of low-energy correlations at low temepratures.

\subsection{Formation mechanism of the heavy fermion state and the excess $\mathrm{Mn}$ effect}
One of the central issues in YMn$_{2+\delta}$Zn$_{20-x}X_x$ is whether the heavy-fermion behavior originates from the intrinsic Mn pyrochlore lattice or from localized moments associated with excess Mn atoms occupying Zn sites. Our site-selective NMR/NQR measurements provide strong evidence to address this issue.
By isolating the signals from the 16$d$ pyrochlore sites, we observed a clear enhancement of $1/T_1T$ at low temperatures, even in the presence of excess Mn impurities.
Furthermore, the value of $1/T_1T$ at the 16$d$ Mn site is larger than that at the In site, suggesting that the 3$d$ electrons of the 16$d$ Mn atoms play a dominant role in the magnetism.
In addition, if the enhancement of the relaxation rate were governed by spin fluctuations from the localized moments of impurities, $1/T_{1}T$ would be expected to follow Curie's law ($\propto 1/T$) rather than becoming constant at low temperatures.
These results strongly suggest that the heavy-fermion state, characterized by a large electronic specific heat coefficient $\gamma \approx 280$ mJ K$^{-2}$ mol$^{-1}$, is intrinsic to the Mn pyrochlore network rather than a spurious effect of localized impurity spins.

However, the mechanism of this mass enhancement is distinct from the conventional antiferromagnetic (AFM) quantum criticality often observed in $f$-electron systems.
According to a typical quantum critical scenario, $1/T_1T$ exhibits a divergent increase as $T \to 0$ due to the development of long-range magnetic correlations\cite{T.Moriya_AP_2000,S.Kitagawa_PRL_2012,S.Kitagawa_JPSJ_2013_B}.
In contrast, $1/T_1T$ in our system is nearly independent of temperature below approximately 1 K.
This Korringa-like behavior indicates the formation of a stable Fermi-liquid ground state after the mass enhancement has occurred.
Furthermore, the absolute magnitude of $1/T_1T$ in the In-substituted system is approximately 20 times smaller than that of YMn$_2$.
In YMn$_2$, strong AFM correlations are driven by the close Mn-Mn distance ($r_\mathrm{Mn-Mn} \approx 2.7$~\AA), placing the system in close proximity to a magnetic instability\cite{Z.Guo-zing_PRB_1999}.
In the YMn$_2$Zn$_{20}$ system, the expansion of the Mn-Mn distance to 5.1 \AA~significantly weakens the exchange interaction $J$.
This suppression of $J$ prevents the development of the strong, nested $q$-dependent fluctuations necessary for quantum criticality.

Instead, it is suggested that the observed heavy-fermion state is more closely related to geometrical frustration inherent to the pyrochlore lattice.
On this corner-sharing tetrahedral network, magnetic order is strongly suppressed, leaving a high density of low-energy magnetic excitations that persist down to low temperatures despite the relatively weak AFM correlations.
Here, the strength of frustration refers not to the absolute magnitude of the interactions, but rather to the degree to which long-range magnetic order is suppressed by the geometrical constraints\cite{A.P.Ramirez_ARMS_1994}.
These frustration-induced fluctuations effectively increase the electron mass without leading to a magnetic phase transition.
The large specific-heat coefficient $\gamma$ in the YMn$_2$Zn$_{20}$ system can be attributed to the development of magnetic correlations and the formation of a coherent band below 10~K.
This is consistent with the general relationship where the product of $\gamma$ and the low-temperature bandwidth is nearly constant.
Consequently, our findings highlight this system as a rare example of a $d$-electron heavy-fermion system where geometrical frustration, rather than conventional AFM quantum critical fluctuations, plays the primary role in mass enhancement.

\section{Conclusion}
Our microscopic investigation of the $d$-electron heavy-fermion candidate $\mathrm{YMn_{2+\delta}Zn_{20-x}In_x}$ and $\mathrm{YMn_{2+\delta}Zn_{20-x}Al_x}$ using NMR and NQR techniques leads to the following conclusions. 
In-substitution selectively occupies the Zn 16$c$ sites, significantly suppressing structural disorder compared with the Al substitution.
The $1/T_1T$ at the Mn site increases at low temperatures, indicating enhanced low-energy excitations associated with the heavy-fermion state.
The site-selective relaxation data further indicate that this enhancement is intrinsic to the Mn pyrochlore network rather than arising from excess Mn moments.
The absolute value of $1/T_1T$ is approximately 20 times smaller than that of YMn$_2$, indicating that the enlarged Mn-Mn distance (5.1 \AA) weakens strong magnetic correlations.
Thus, our results point to a frustration-driven heavy-fermion state in a $d$-electron pyrochlore system, rather than one governed by conventional AFM quantum criticality.
Future studies under external pressure, which can tune the Mn-Mn interactions, will be important for clarifying whether enhanced magnetic correlations can induce novel ground states such as superconductivity.

\section*{Acknowledgments}
This work was supported by Grants-in-Aid for Scientific Research (KAKENHI Grant No. JP20KK0061, No. JP20H00130, No. JP21K18600, No. JP22H04933, No. JP22H01168, No. JP23H01124, No. JP23K22439, No. JP23K25821, and No. JP25H00609) from the Japan Society for the Promotion of Science,  by JST ASPIRE (Grant No. JPMJAP2314), by research support funding from the Kyoto University Foundation, by ISHIZUE 2024 of Kyoto University Research Development Program, by Murata Science and Education Foundation, and by the JGC-S Scholarship Foundation.
Liquid helium is supplied by the Low Temperature and Materials Sciences Division, Agency for Health, Safety and Environment, Kyoto University.

\end{document}